\documentclass[12pt,twoside]{article}   % Specifies the document style.

\usepackage{latexsym} 
\usepackage{mathptm}
\usepackage{epsfig}
\usepackage{times}

\def\beq{\begin{equation}}
\def\u{\underline}
\def\eeq{\end{equation}}
\def\beqn{\begin{eqnarray}}
\def\eeqn{\end{eqnarray}}
\def\slash{/\hspace*{-2.5mm}}
 
\date{} 
\begin{document}

\markboth{SABINE HOSSENFELDER}{Cosmological Consequences of Anti-gravitation}

\title{Cosmological Consequences of Anti-gravitation}
\author{S.~Hossenfelder\thanks{sabine@physics.ucsb.edu}\\
\vspace*{-.2cm}\small{Department of Physics,University of California}\\ 
\vspace*{-.2cm}\small{Santa Barbara, CA 93106-9530, USA}}
  
\date{}
\maketitle

\begin{abstract} 
The dynamics of a universe with an anti-gravitating contribution to the matter content is examined. The
modified Friedmann equations are derived, and it is shown that anti-gravitating radiation is the 
slowest component to dilute when the universe expands. Assuming an interaction between both kinds of matter
which becomes important at Planckian densities, it is found that the universe undergoes a periodic cycle of
contraction and expansion. Furthermore, the possibility of energy loss in our universe through separation of both types of
matter is discussed. 
\end{abstract}

\maketitle

\section{Introduction}
  
During the last decades, experimental achievements in astrophysics provided us with new insights about
our universe. The more precise our observations have become, the more obvious also the insufficiencies of
our understanding have become. Today's research in cosmology is accompanied by the group of
cosmological problems, which strongly indicate that our knowledge about the universe is incomplete.

In a previous work \cite{Hossenfelder:2005gu}, a framework has been introduced to 
include negative gravitational sources into General Relativity ({\sc GR}). Here, it is examined how these influence 
the evolution of the universe, which can provide a useful basis to address some of the cosmological problems.

The proposed introduction of an anti-gravitating sector is a relaxation of
the equivalence principle. The new 
anti-gravitating particles are defined through their transformation
behavior under general coordinate transformations, in which they differ from the usual particles. 
This leads to the introduction of a modified covariant derivate, followed by a modified equation of
motion.

The so defined anti-gravitating matter has a negative gravitational stress-energy, even 
though its kinetic energy remains positive. This theory therefore does not suffer from  
instabilities through decaying vacuum fluctuations. The gravitational energy 
acts as a charge, which can take either positive or negative sign. Since the interaction between
both types of particles is mediated by gravity only, it is suppressed by the large value of the Planck scale,
and naturally very weak at the present day.

The proposal of a gravitational
charge symmetry has previously been examined in Refs. \cite{linde1,linde2,Moffat:2005ip,Kaplan:2005rr}. 
Furthermore, there have been various approaches \cite{Bondi,Quiros:2004ge,Borde:2001fk,Davies:2002bg,Ray:2002ts,
Rosenberg:2000cv,Torres:1998cu,Zhuravlev:2004vd,Faraoni:2004is,Henry-Couannier:2004vt,Nickner:2006xh} 
to include anti-gravitating matter into quantum 
theories as well as into {\sc GR}. Also, the topic of negative energies has recently received attention within the context
of  ghost condensates \cite{Arkani-Hamed:2003uz,Arkani-Hamed:2003uy,Arkani-Hamed:2005gu,Mann:2005jz,Krause:2004bu}. 

This paper is organized as follows: In the next section, the properties of the anti-gravitating
matter are briefly recalled. In section \ref{hub}, we solve the geodesic equations for both
types of particles in the Friedmann-Robertson-Walker background, and derive Hubble's law. 
In section \ref{evolution} we
examine the evolution of the universe. After a brief
discussion in section \ref{dis}, we conclude in section \ref{concl}.

Throughout this paper we use the convention $c=\hbar=1$. The signature of the metric is $(-1,1,1,1)$. Small
Greek indices $\kappa,\nu,\epsilon...$ are space-time indices.
  
\section{Definition of Anti-Gravitating Matter}

The unitary representations $U$  of a gauge group $G$ define the transformation behavior of particle fields $\Psi$. For
two elements of the group $g,g'$, the representation fulfils
\beqn
U(g) U(g') = U(gg') \label{multrep}\quad,
\eeqn
and the field transforms as $\Psi \to \Psi' = U(g) \Psi$.
From this, it is always possible to construct a second representation, defined by 
\beqn
\widetilde{U}(g) = (U(g^{-1}))^T \quad,
\eeqn
which belongs to the charge-conjugated particle. The anti-particle $\overline{\Psi}$
transforms according to the contragredient representation, ${\overline{U}}$, which is
${\overline U}(g) = U(g^{-1})$.   

In case of a local symmetry, these transformations lead to the introduction of
gauge-covariant derivatives in the usual way. Suitable combinations of particles with anti-particles allow
to construct gauge-invariant Lagrangians.

\subsection{Transformation Properties}

From the above, one is tempted to conclude that there is no charge-conjugation for gravity. If the gauge-group is the 
Lorentz-group $SO(3,1)$, then the elements $\Lambda$ fulfill
$\Lambda^{-1} = \Lambda^T$,
which means that in this case the second representation $\widetilde{U}$ is equivalent to $U$. 

However, this does not apply when the field transforms under a general coordinate transformations $G$. Let $\Psi$ be
a vector field and an element of the tangent space $TM$. Under a general coordinate transformation $G$, the field and
its conjugate behaves as
\beqn
TM~ &:& \Psi \to \Psi' =  G \Psi \quad,\quad 
{TM^*} :  {\overline \Psi} \to {\overline \Psi'} = {\overline \Psi} G^{-1} \quad, \label{stdtrafo}
\eeqn
where $TM^*$ is the dual to $TM$. 

The equivalence principle requires that the fields in the tangential space 
transform like in Special Relativity. I.e. if $G$ is an element of the Lorentz-group, the fields have
to transform like Lorentz-vectors. However, the generalization to a general coordinate transformation is
not unique. Instead of Eq. (\ref{stdtrafo}) one could have chosen the field to transform according to
\beqn
{\underline{TM}}~ : {\underline\Psi} \to {\underline\Psi}' = (G^T)^{-1} \underline\Psi  \quad,\quad
{\underline{TM}}^* &:&   {\underline{\overline \Psi}} \to {\underline{\overline \Psi'}} = 
{\underline{\overline \Psi}} G^T \quad. \label{trafo}
\eeqn
Here, the space ${\underline{TM}}$ is a vector-space which spans the basis for these fields, and ${\underline{TM}}^*$ is
its dual. In case $G$ was an element of the Lorentz-group, i.e. $G^{-1}=G^T$, both representations 
(\ref{stdtrafo},\ref{trafo})
agree. For general coordinate transformations that will not be the case. Indeed, one sees in particular that 
the newly introduced fields will have a modified scaling behavior. 
 
For the following, it is convenient to introduce a map $\tau$ which, in the vector-representation, is a 
vector-space isomorphism from $TM$ to ${\underline{TM}}$. For the map $\tau: \underline\Psi = \tau \Psi$ 
to transform adequately, $\underline\Psi' = \tau' \Psi'$, one finds the behavior
\beqn
\tau' = (G^T)^{-1} \tau G^{-1} \label{trafotau} \quad.
\eeqn

\begin{figure}[t]
\vspace*{-0.0cm}
\centering \epsfig{figure=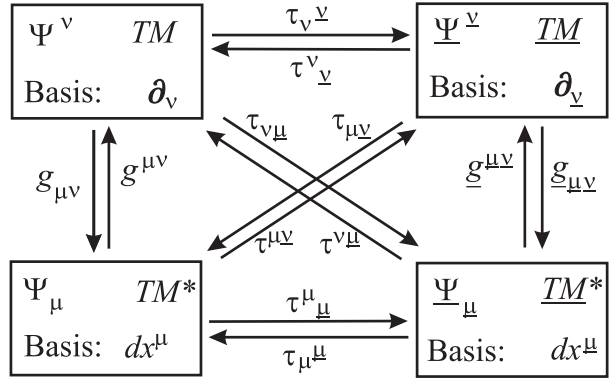, width=8cm}
 
\caption{Relations between the maps. The left side depicts 
the  tangential and co-tangential spaces. The right side depicts the corresponding
spaces for the anti-gravitational fields.\label{fig1}}
\end{figure}

It will be useful to clarify the emerging picture of space-time properties by looking at 
a contravariant vector-field $\Psi^{\kappa}$ as depicted in Figure \ref{fig1}. This field 
is a cut in the tangent bundle, that is the set of tangent spaces $TM$ at every point of 
the manifold which describes our space-time.
The field is mapped to its covariant field, $\Psi_{\nu}$, a cut in the co-tangent bundle, $TM^*$, by the 
metric tensor $\Psi_{\nu} = g_{\kappa \nu} \Psi^{\nu}$. 

%#######################################################################

%#######################################################################

The newly introduced field  
${\underline\Psi}^{\underline\kappa}$ (from here on named anti-gravitating)
transforms in the tangential space like a Lorentz-vector in Special Relativity. But it 
differs in its behavior under general coordinate transformations. 

In order not to spoil the advantages of the Ricci-calculus, it will be useful to introduce a basis
for the new fields that transforms accordingly.  Locally, the new field can be expanded in this 
basis ${\partial}_{\underline\kappa}$. These basis elements form
again a bundle on the manifold, that is denoted with ${\underline{TM}}$. To each of the elements of
 ${\underline{TM}}$ also a dual space exists, defined as the space of all linear maps on ${\underline{TM}}$. This
 space is denoted by ${\underline{TM}}^*$ and its basis as $d{{x}^{\underline\kappa}}$.
The map from ${\underline{TM}}$ to ${\underline{TM}}^*$ will be denoted 
${\underline g}_{{\underline\kappa}{\underline\nu}}$, and defines a scalar product on ${\underline{TM}}$. 
The relation between the introduced quantities is summarized in Figure \ref{fig1}.  

Note, that the underlined indices on these quantities do not
refer to the coordinates of the manifold but to the local basis in the tangential spaces. All of these
fields still are functions of the space-time coordinates $x_\nu$. Also, the map 
$\underline g_{\underline \kappa\underline\nu}$, is not a metric on the manifold, 
and does not measure physical distances and angles. Instead, it is a scalar product on the underlined spaces.
 
The transformation behavior in Eq. (\ref{trafotau}) together with the observation 
that in a local orthonormal basis both fields transform identical under 
local Lorentz-trans\-formations, 
gives us an explicit way to construct $\tau$. We choose a local orthonormal basis $\hat e$ in $TM$, 
which is related to the coordinate basis by the locally linear map $E \hat e = \partial$. In this basis, 
the metric is just $\eta$ and $\hat \tau$ is just the identity. One then finds $\tau$ in a general coordinate
system by applying Eq.(\ref{trafotau})
\beqn
\tau = (E^T)^{-1} \hat \tau E^{-1} = (E E^T)^{-1}  \quad \label{tauE}.
\eeqn
An example for this is given in section \ref{example}.

\begin{figure}[t]
\vspace*{-0.0cm}
\centering \epsfig{figure=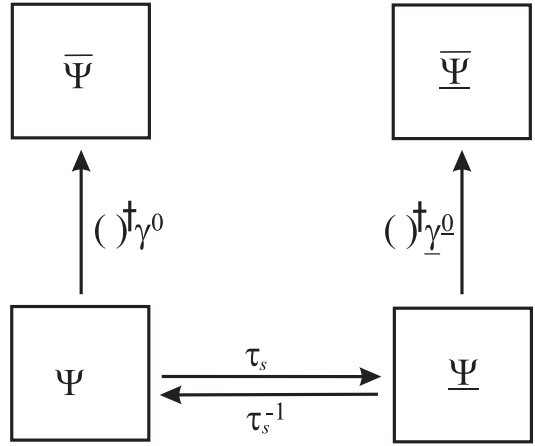, width=7cm}
 
\caption{Relations between the maps for fermionic fields.\label{fig2}}
\end{figure}

The properties of the vector-fields are transferred directly to those of fermionic fields by using the fermionic
representation and transformations. In this case, 
the map $(\;)^\dag \gamma^0$, instead of the metric, is used to relate a particle to the particle transforming under 
the contragredient representation, and the map relating the spinor-bundles is denoted with $\tau_s$ (spinor
indices are suppressed), see Fig. \ref{fig2}. With this notation it 
is $\underline \Psi = \tau_s \Psi$.

%#######################################################################

%#######################################################################

\subsection{The Covariant Derivative}

It is now straightforward to introduce a covariant derivative for the new fields, in much the same way 
as one usually introduces the derivative for the charge-conjugated particles. 
We will use the notation $\nabla$ for the general-covariant derivative and
$D$ for the general covariant derivative including the gauge-derivative of the fields. It is understood
that the form of the derivative is defined by the field it acts on, even though this will not
be noted explicitly. 

For an element $V_\nu$ of $TM^*$, besides the usual derivative $\nabla_\alpha V_\nu$, one can now apply the
composite derivative $\tau_\nu^{\;\; \underline \kappa} \nabla_\alpha \tau^\kappa_{\;\; \underline \kappa} V_\kappa$, 
or, using a more abstract notation $\tau \nabla \tau^{-1}$. Similar considerations apply for elements of products of
$TM^*$ and $TM$, or spinor bundles, with appropriate applications of the map $\tau$, or $\tau_s$ respectively. 

One introduces the derivative in the direction of $\nu$ on the basis in a 
general way by
\beqn
\nabla_\nu \partial_{\kappa} &=& \Gamma^{\epsilon}_{\; \nu \kappa} ~ \partial_\epsilon  \quad,\quad
\tau_\kappa^{\;\; \underline \kappa} {\nabla}_{\nu} \tau^\alpha_{\;\; \underline \alpha }{\partial}_{\alpha} = 
\left( \tau_\kappa^{\;\; \underline \kappa} ~\tau^{\epsilon}_{\;\; \underline \epsilon}
\Gamma^{{\underline\epsilon}}_{\;\; \nu {\underline\kappa}}\right) {\partial}_{\epsilon} \quad. \label{stdder}
\eeqn
Note that the connection symbols do not transform homogeneously in the first and third index and therefore these 
indices can not be contracted with the $\tau$'s. Alternatively, one can define the connection coefficients with respect to
$\tau \nabla \tau^{-1}$ 
\beqn
{\underline \Gamma}^\epsilon_{\;\;\nu\kappa} =
\tau_\kappa^{\;\; \underline \kappa} ~\tau^{\epsilon}_{\;\; \underline \epsilon}
\Gamma^{{\underline\epsilon}}_{\;\; \nu {\underline\kappa}} \quad. \label{ugamma}
\eeqn
The operators $\tau \nabla \tau^{-1}$, and $\tau \partial \tau^{-1}$ fulfill the same commutation relations as 
$\nabla$ and $\partial$, which means in particular that $\underline \Gamma$ is symmetric in the lower two indices.
To get an explicit formula for these Christoffelsymbols one commonly uses the requirement that the
covariant derivative on the metric itself vanishes  
$
\nabla_\lambda g_{\nu \kappa} =0$ which assures that the scalar product is covariantly conserved. Similarly, 
we require $\tau_\nu^{\;\; \underline \nu}\tau_\kappa^{\;\; \underline \kappa}~ 
\nabla_{\lambda} \underline g_{\underline\nu \underline\kappa} = 0$.
  
From this one finds \cite{Hossenfelder:2005gu} the explicit expression
\beqn
{\underline \Gamma}^{\nu}_{\;\; \lambda \kappa} &=& 
\frac{1}{2} {g}^{\nu \alpha} 
\left( \tau_{\kappa}^{\;\;\underline \kappa}\tau_{\alpha}^{\;\;\underline \alpha}
\partial_{\lambda} {\underline g}_{\underline{\kappa\alpha}} +  
\tau_{\lambda}^{\;\;\underline\lambda} \tau_{\alpha}^{\;\;\underline\alpha} \partial_{{\kappa}} 
{\underline g}_{\underline{\lambda \alpha}} 
- 
\tau_{\kappa}^{\;\;\underline\kappa} \tau_{\lambda}^{\;\;\underline\lambda}
\partial_{\alpha} {\underline g}_{\underline\kappa \underline\lambda} \right) \label{chrisagu} \quad,
\eeqn
or, with Eq.(\ref{ugamma})
\beqn
\Gamma^{\underline\nu}_{\;\; \lambda \underline\kappa} &=& \frac{1}{2} {\underline g}^{\underline\nu \underline\alpha} 
\left( \partial_{\lambda} {\underline g}_{\underline{\kappa\alpha}} +  
\tau_{\lambda}^{\;\;\underline\lambda} \tau^{\kappa}_{\;\;\underline\kappa} \partial_{{\kappa}} 
{\underline g}_{\underline{\lambda \alpha}} 
- 
\tau_{\lambda}^{\;\;\underline\lambda} \tau^{\alpha}_{\;\;\underline\alpha}
\partial_{\alpha} {\underline g}_{\underline\kappa \underline\lambda} \right) \label{chrisag} \quad.
\eeqn
The use of the tetrad $g_{\mu\nu} = \eta_{ij}E^i_{\;\; \mu} E^j_{\;\;\nu}$,
together with 
Eq.(\ref{tauE}), the
definition $\underline g_{\underline \kappa \underline \nu}=
\tau^\kappa_{\;\; \underline \kappa}\tau^\nu_{\;\;\underline \nu}~ g_{\kappa\nu}$, and the above Eq.(\ref{chrisag})
provides an explicit construction for the dynamics of the new fields. 

With these generalized covariant derivatives, one obtains the Lagrangian of the anti-gravitational 
field by replacing all quantities with the corresponding anti-gra\-vi\-tational quantities
and using the appropriate derivative for the new fields to assure homogenous transformation behavior. 
One then finds additional possibilities to construct gauge- and Lorentz-invariant kinetic 
terms in the Lagrangian. E.g. for fermionic fields $\Psi$, besides
the usual $\overline \Psi \slash D \Psi$ one can now also have $\overline \Psi \tau_s^{-1} \slash D \tau_s \Psi 
= \overline{\underline \Psi} \slash D \underline \Psi$, which is not identical to the first because the covariant 
derivative on $\tau_s$ is (in general) non-vanishing. In case one considers operators with higher order derivatives, more
choices become possible (see also section \ref{early}).

\subsection{Geodesic Motion}

It is instructive to look at the motion of a classical test particle by considering  
the analogue of parallel-transporting the tangent vector. The particle's world line is
denoted ${x}_\nu (\lambda)$, and the anti-gravitating particle's world line is denoted 
${\underline x}_\nu (\lambda)$\footnote{A word of caution is necessary for this notation: the underlined
${\underline x}_\nu$ does only indicate that the curve belongs to the anti-gravitating particle; it is
not related to the curve of the usual particle ${x}_\nu$.}. 
   
In contrast to the gravitating particle,
the anti-gravitating particle parallel transports not its tangent vector ${\underline t}^\alpha = d{\underline x}^{ \alpha} /d\lambda$, 
but instead the related quantity in ${\underline {TM}}$,  which corresponds to the kinetic momentum, and is
${\underline t}^{\underline \alpha}=  \tau_\alpha^{\;\;\underline \alpha} {\underline t}^{\alpha}$.  On
the particle's world line ${\underline x}_\nu (\lambda)$, it is ${\underline t}^{\underline \alpha}$ which
is covariantly conserved. 
Parallel transporting is then expressed in evaluating the derivative in direction 
of the curve and set it to zero. For the
usual geodesic, which parallel transports the tangential vector, one has
$t^\nu  
{\nabla}_{\nu}~ {t}^{\alpha} = 0  
$, 
whereas for the anti-gravitating particle one has
\beqn
{\underline t}^{\nu} ~\tau^\alpha_{\;\; \underline\alpha}~ 
{\nabla}_{\nu}~ \tau_\kappa^{\;\; \underline \alpha}~ {\underline t}^\kappa = 0\quad, \label{agpt}
\eeqn
which agrees with the usual equation if and only if the covariant derivative 
on $\tau_\kappa^{\;\; \underline \alpha}$ vanishes\footnote{One might wonder, whether these two curves for 
$t^\nu$ and $\underline t^\nu$ are equal up to a mere re-parametrization $d \lambda \to \tilde d \lambda$ of the curve. In such
a case $t^\nu \to \tilde t^\nu = d\lambda/d\tilde \lambda t^\nu$ which could only corresponds to 
$\tau^\nu_{\;\;\underline \nu}$ being some scalar function times the identity. Such a scalar however, is fixed by
the requirement that in a local orthonormal basis $\tau$ is the identity, and therefore does not cause a reparemetrization. 
We will also see later that $\tau$ is not in general proportional to the identity matrix.}. It is important to note that the tangent 
vector ${\underline t}^\alpha$ is not parallel transported along the curve given by the new Eq. (\ref{agpt}).

Using  the covariant derivative  
${\nabla}_{\nu} {\underline t}^{\underline \alpha} = 
{\partial}_{\nu} {\underline t}^{\underline \alpha}
+   
\Gamma^{{\underline \alpha}}_{\;\; {\nu}{\underline \epsilon}}~ 
{\underline t}^{\underline \epsilon}$, 
  one obtains
\beqn
 \frac{d {\underline x}^{\nu}}{d \lambda} 
\cdot \left( \partial_{\nu} \frac{d {\underline x}^{{\underline \alpha}}}{d \lambda} 
  +  
\Gamma^{\underline \alpha}_{\;\; \nu \underline \epsilon } \frac{d {\underline x}^{\underline \epsilon}}{d \lambda} 
\right) = 0 \quad,
\eeqn
and by rewriting $\partial_{\nu}= (d \lambda /d {\underline x}^{\nu})~ d/d\lambda$ one 
 finds the anti-geodesic equation
\begin{eqnarray} \label{Antigeo}
\frac{d^2 {\underline x}^{{\underline \alpha}}}{d\lambda^2} + \tau^\nu_{\;\;\underline \nu}
\Gamma^{\underline \alpha}_{\;\; \nu \underline \epsilon}
\frac{d {\underline x}^{{\underline \epsilon}}}{d\lambda}
\frac{d {\underline x}^{{\underline \nu}}}{d \lambda} =0 \quad.
\end{eqnarray} 
This equation should be read as an equation for the quantity ${\underline t}^{\underline \alpha}$ rather than an equation
for the curve. To obtain the curve, one proceeds as follows
\begin{itemize}
\item Integrate Eq. (\ref{Antigeo}) once to obtain ${\underline t}^{\underline \alpha}$,
\item Translate this into the geometric tangential vector ${\underline t}^{\underline \alpha}=  \tau_\alpha^{\;\;\underline \alpha} {\underline t}^\alpha$,
\item Integrate a second time with appropriate initial conditions to obtain ${\underline x}^{\alpha}$.
\end{itemize}

It is important to note that the equations of motion Eq.(\ref{Antigeo}) are invariant under general
diffeomorphism, provided that the quantities are transformed appropriately.

\subsection{The Equivalence Principle}

For a particle moving in a curved spacetime, it is possible to choose a freely falling coordinate system, in which
the usual Christoffelsymbols vanish. However, this freely falling frame for the particle will in general
not also be a freely falling frame for the anti-gravitational particle. Consequently, the Christoffelsymbols in 
Eq.(\ref{chrisag}) will in general not also vanish in the freely falling frame of the usually gravitating particle. 
Both sets of symbols therefore will in general not be proportional to each other\footnote{In a globally flat spacetime,
it is possible to choose $g=\eta$. It is then also ${\underline g} =\eta$, and $\tau =$ Id. 
In these coordinates, the derivative on $\tau$ vanishes and
the geodesic and the anti-geodesic curve agree. Since both curves are invariant under coordinate transformations, they will
agree for every choice of coordinates in a globally flat spacetime.}.

This explains how it is possible to have a relaxation of the equivalence principle. Locally, both particles
experience an gravitational downwards pull like an acceleration in flat space. 
For the usual particle, inertial mass equals gravitational mass, and upward
acceleration corresponds to a downward attractive gravitational pull. 
For the anti-gravitating particle, the inertial mass is the negative of the gravitational mass, and upward
acceleration corresponds to a downward repulsive gravitational push. The first is realized for particles obeying the usual transformation
properties, the second for particles obeying the transformation properties Eq.(\ref{trafo}). 

A scenario with negative gravitational masses is commonly believed to lead to inconsistencies. When neglecting
an appropriate covariant derivative, a particle with negative gravitational mass must move on a geodesic in the
field of a positively gravitating mass, since the curve is independent on the particle's mass. 
On the other hand, a  particle with positive gravitational mass in the field of a negatively gravitating body is repelled, 
as one most easily concludes using the Newtonian limit. Taken together, the positive mass will attract the negative one,
whereas the negative mass will repel the positive one. This - in the limit of two point particles -
leads to a self-accelerating system. Such considerations go back to \cite{Bondi}, and can be resolved by noting that
the curve of the negatively gravitating mass in the positively gravitating background is not a 
geodesic. In this case, like charges attract, unlike charges
repel, and a self-acceleration can not occur. 

Even though both particles on their own can not distinguish between acceleration in flat space and local
gravitational effects, a pair of both can. In flat space, accelerating a pair of gravitating and anti-gravitating masses 
will not allow to distinguish between both as only the inertial masses are involved. In a gravitational field 
however, the one will move in the opposite direction from the other. Therefore, the equivalence principle is
not generalized, but relaxed: it holds for both particles separately, but not for both together when they can compare their
ratios of inertial to gravitational mass.

\subsection{The Stress-Energy Tensors}
    
One finds  the gravitational Stress-Energy Tensors ({\sc SET}s) for the fields from variation of the action
with respect to
the metric
\beqn
S &=& \int {\rm d}^{d+4}x~ \sqrt{-g} \left[   G~{\mathcal R} 
+  {\mathcal L} + {\underline {\mathcal L}}
\right] 
\label{full} \quad,
\eeqn
where $G=1/m_p^2$, ${\mathcal R}$ is the curvature scalar, ${\mathcal L}$ is the Lagrangian of usual fields, and ${\underline {\mathcal L}}$ is the
Lagrangian of the new fields.  The source terms then take the form \cite{Hossenfelder:2005gu}
\beqn
 T^{\kappa \nu} &=&   \frac{\delta {\mathcal L}}{\delta {g}_{\kappa \nu}} - \frac{1}{2} g^{\kappa \nu} {\mathcal L} \quad,\\
{\underline T}^{\kappa \nu} &=&  - \tau^\kappa_{\;\; {\underline \kappa}}  \tau^\nu_{\;\; {\underline \nu}} 
\left( \frac{\delta {\underline {\mathcal L}}}{\delta {\underline g}_{{\underline \kappa} {\underline \nu}}} + \frac{1}{2} g^{{\underline \kappa} {\underline \nu}} 
{\underline {\mathcal L}} \right) \label{sign} \quad.
\eeqn
Under a perturbation
of the metric, the anti-gravitational fields will undergo a transformation exactly opposite to these of the normal
fields as one expects by construction. The $\tau$-functions convert the indices and the transformation behavior 
from ${\underline{TM}}$ to the usual tangential space.

The Lagrangian Eq.(\ref{full}) consists of the gravitational contribution and two types of matter. In the
following it is assumed that the dominant type of matter in our universe is the positively gravitating one. 
%\footnote{Indeed, if this
%was not the case, the vacuum should be defined with respect to $-\eta$ instead of $\eta$. The field equations
%remain invariant under subsequent exchange of gravitating with anti-gravitating fields and an overall 
%sign change in the metric which has also been pointed out in \cite{Quiros:2004ge}. 
%The $-\eta$ background can 
%be interpreted as a second vacuum ground state.}. 
We will not consider the
question of a transition of one dominance to the other. As will become apparent later, this can not occur in
the here examined case.

If one considers the Lagrangian of the Dirac-field  one has to formulate the action in form of the tetrad fields.
The above used 
argument then directly transfers to the Dirac field through the properties of the anti-gravitational field under
diffeomorphisms. For the fermions, the {\sc SET} then can be simplified inserting
that the field fulfills the equations of motion $D\hspace*{-.25cm}/\hspace*{.15cm} \Psi = D\hspace*{-.25cm}/\hspace*{.15cm} 
\underline \Psi =0 $. 

Most importantly, one sees that the {\sc SET} of the anti-gravitating field yields a contribution
to the source of the field equations with a minus sign (and thus justifies the name anti-gravitation). This 
is due to the modified transformation behavior of the field components. However, we also see that the second term, 
arising from the variation of the metric
determinant, does not change sign. Fermionic matter will therefore display different properties than radiation. 
This will turn out to be an important ingredient to the evolution of these fields in an expanding universe. 

The corresponding conservation law of the derived source terms which follows from the Bianchi-identities is as
usual $\nabla^\nu ( T_{\kappa \nu} + {\underline T}_{\kappa \nu}) = 0$.
 
It is crucial to note that the kinetic {\sc SET} as defined from the Noether current does {not} have
a change in sign. Here, no variation of the metric is involved and the gravitational properties of the fields
do not play a role. To clearly distinguish this canonical {\sc SET} from the gravitational
source term,  the canonical {\sc SET} is denoted with $\Theta_{\nu \kappa}$, whereas  
the above used $T_{\nu \kappa}$ is kept for the gravitational {\sc SET}.

The canonical {\sc SET} for a matter Lagrangian
${\mathcal L}(\Psi, \nabla_\nu \Psi)$   from Noether's theorem is 
\beqn
\Theta^{\nu}_{\;\; \kappa} = \frac{\partial {\mathcal L}}{\partial (\nabla_\nu \Psi) } \nabla
_\kappa \Psi - \delta^{\nu}_{\;\; \kappa} {\mathcal L} \quad,
\eeqn
and is covariantly conserved $\nabla_\nu \Theta^{\nu}_{\;\;\kappa} = 0$. 
Correspondingly, one finds the conserved current for the anti-gravitational field
\beqn
{\underline \Theta}^{\nu}_{\;\; \underline \kappa} = 
\tau^{\kappa}_{\;\;\underline \kappa} 
\left( \frac{\partial {\mathcal {\underline L}}}{\partial ({\nabla}_\nu {\underline \Psi}) } 
{\nabla}_{\kappa} {\underline \Psi} - \delta^{\nu}_{\;\;\kappa} 
{\mathcal {\underline L}} \right) \quad,
\eeqn
which is also covariantly conserved $\nabla_\nu \underline \Theta^{\nu}_{\;\;\underline \kappa} = 0$. 

However, in general the quantities derived from Noether's theorem are neither symmetric, nor are 
they traceless or gauge-invariant. For {\sc GR}, Belinfante's symmetrized tensor is the more adequate 
one \cite{Babak:1999dc,enmomt1,enmomt2} which we will also use in the following.

From the 
Noether current, one gets a  total conserved quantity for each space-time direction $\kappa$, which
obey the conservation equations
\beqn
\nabla^{\nu}  \Theta_{\nu \kappa} + \tau_\kappa^{\;\; \underline \kappa}~ 
 {\nabla}^{\nu} {\underline \Theta}_{\nu \underline\kappa} = 0 \quad. \label{cons2}
\eeqn 
 
The form of the second term of Eq. (\ref{cons2}) is readily interpreted: when the anti-gravitating 
particle gains kinetic energy 
on a world line, the gravitational particle would loose energy when traveling on the same world line. 
The interaction with the gravitational field is inverted.  
 
One thus can identify the anti-gravitating particle as a particle whose kinetic momentum vector transforms
under general diffeomorphism according to Eq.(\ref{trafo}), whereas the standard particle's kinetic 
momentum transforms according to Eq.(\ref{stdtrafo}).

In the present low energy epoch of the universes evolution, the interaction between gravitating
and anti-gravitating matter is very weak. Since both types of matter repel, it is natural
to expect that we live today in a sector of the universe with predominantly one type of matter.
In this case, one can neglect the presence of anti-gravitating matter for most purposes and the
here proposed framework reduces to the standard {\sc GR}. However, the presence of anti-gravitating
matter might be relevant at large distances, at high densities, and in strongly curved backgrounds.

\subsection{Example: Newtonian Limit in Schwarzschild Background}
\label{example}

To give an example, let us consider the familiar case of a particle moving in a Schwarzschild-metric with the
line-element
\beqn
ds^2 = - \gamma dt^2 + \frac{1}{\gamma} dr^2 + r^2 ( d\theta^2 + \sin^2 \theta d \phi^2) \quad,
\eeqn           
with $\gamma = 1 - 2M/r$. Going to a local orthonormal basis yields 
\beqn
E^i_{\;\; \nu} = (E^i_{\;\; \nu})^T = {\rm{diag}}(\sqrt{\gamma},\sqrt{\gamma},1/r,1/(r \sin \theta) ) \quad,
\eeqn
and so
\beqn
\tau^\nu_{\;\; \underline \nu} &=& {\rm{diag}} (\frac{1}{\gamma},\gamma,1/r^2,1/(r^2 \sin^2 \theta)) \quad,\label{tausm}\\
\tau_{\nu \underline \nu} &=& {\rm{diag}} (1,1,1,1) \quad,\\
\underline g_{\underline \nu \underline \kappa} &=& {\rm{diag}} (- \frac{1}{\gamma},\gamma,1/r^2,1/(r^2 \sin^2 \theta)) \quad.
\eeqn
We will consider a radially moving particle with ${\underline t}^\phi = {\underline t}^\theta = 0$. Since $\tau$ is diagonal, we also
have ${\u t}^{\underline \phi} = {\u t}^{\underline \theta} = 0$. 
Assuming a small velocity of the test-particle $v/c \ll 1$, and $t \approx \lambda$, one approximates the geodesic equation 
as usual to
\beqn
\dot t^{\underline r} \approx - 
\tau^t_{\;\; \underline t} \Gamma^{\underline r}_{\;\; t {\underline t}} \quad, 
\eeqn
where a dot denotes the derivative with respect to $t$. With Eq.(\ref{chrisag}) one computes the Christoffelsymbols 
of the anti-gravitating field (see Appendix A). 
Using these, one finds to first order   in the Newtonian limit the expression 
\beqn
\dot {\underline t}^{\underline r} \approx  - \frac{1}{2}
\partial_r \underline g_{\underline t \underline t} = \frac{M}{r^2} + {\mathcal{O}}(r^{-3}) \quad, 
\eeqn 
Integrating once and converting the index results in the approximation 
\beqn
\dot r &\approx& - \frac{M}{r} + {\mathcal{O}}(r^{-2}) + {\rm const.} \quad, \\
\ddot r &\approx&  \frac{M}{r^2} + {\mathcal{O}}(r^{-3}) \quad.
\eeqn
One sees that indeed the anti-gravitating particle is repelled by the gravitational mass of the
background field.

\section{Hubble's law}
\label{hub}

We live today in a galaxy in which one gravitational type of matter is dominant. This local
dominance, however, must not be identical to the global dominance. Though  we have defined 
the globally dominating matter to be the 'positive' one, it is not {\sl a priori} clear 
whether our galaxy is of the same type of matter. It is in principle possible that we are made of 
anti-gravitating matter, which, in a local surrounding of the same matter type, 
would be indistinguishable from being positively gravitational matter in a local surrounding 
of positively gravitating matter. Whether we are of the globally dominating matter type or not,
one would expect this to reflect in cosmological observations. 

From
analyzing the equations of motions of test particles as well as those of the cosmic fluid, we will in
the following see that we are indeed of the dominating positively gravitating type of matter.

Starting point is the usual assumption of an isotropic and homogeneous universe which leads to the
general line-element of the Friedmann-Robertson-Walker metric:
\beqn
ds^2 = -dt^2 + a^2(t)\left[dr^2 + r^2 \left( d\theta^2 + \sin^2 \theta d \phi^2 \right) 
\right] \quad. \label{FRW}
\eeqn 
The Christoffel symbols of this geometry for both types of particles are given in Appendix B.
Going to a local orthonormal basis one finds 
\beqn
E^i_{\;\; \nu} = (E^i_{\;\; \nu})^T = {\rm{diag}}(1,\frac{1}{a},\frac{1}{ar},\frac{1}{a r \sin \theta} )\quad,
\eeqn
and so
\beqn
\tau^\nu_{\;\; \underline\nu} &=& {\rm{diag}} (1,\frac{1}{a^2},\frac{1}{a^2 r^2},\frac{1}{a^2 r^2 \sin^2 \theta}) \quad, 
\label{taufrw}\\
\tau_{\nu \underline\nu} &=& {\rm{diag}} (1,1,1,1) \quad,\\
\underline g_{\underline\nu \underline\kappa} &=& {\rm{diag}} (- 1,\frac{1}{a^2},\frac{1}{a^2 r^2},\frac{1}{r^2 \sin^2 \theta}) \quad.
\eeqn

We will now examine the motion of the anti-gravitating photon. 
For a clear analysis, this is done parallel to the standard case.  
We will denote the momentum of the gravitating particle with
$t_\nu$ and that of the  anti-gravitating with $\u{t}_\nu$, in accordance with the
notation of the previous section. The curves differ in the fact that the one transports the usual tangential vector,
whereas the other transports the vector which is subject to the new transformation behavior.  

By assuming a radial motion one can simplify the equations of motion with
$t^\theta = t^\phi= {\u t}^\theta  =  {\u t}^\phi= 0$. With use of Appendix B,
the geodesic equations read
\beqn
\frac{d }{d \lambda} t^t&=& - a \dot a \left( t^r \right)^2  \label{stdgeo1} \\
\frac{d   }{d \lambda} t^r &=& - 2 \frac{\dot a}{a} t^r t^t \quad, \label{stdgeo2}
\eeqn
whereas the anti-geodesic equations take the form 
\beqn
\frac{d  }{d \lambda} {\u t}^{\u t}  &=&  \frac{\dot a}{a^3} \left( {\u t}^{\u r} \right)^2  \label{newgeo1} \\
\frac{d  }{d \lambda} {\u t}^{\u r}  &=& 2 \frac{\dot a}{a} {\u t}^{\u r} 
{\u t}^{\u t} \label{newgeo2} \quad.
\eeqn
One solves Eqs.(\ref{stdgeo1}),(\ref{stdgeo2}) with 
\beqn
\omega:= t^t = \frac{c_i}{a(t)} \quad, \quad t^r = \frac{c_i}{a(t)^2} \quad,  
\eeqn
where $c_i$ is some constant initial value. To solve the equations of motion of the anti-gravitating particle, 
one makes an {\sl ansatz} as powerlaw for the kinetic energy  
${\underline\omega}:={\u t}^{\u t} = c_i a^n$, 
which yields in Eq.(\ref{newgeo1})
\beqn
{\u t}^{\u r} = c_i \sqrt{n}~ a^{n+1} \quad,
\eeqn
where we have used that $d t = \tau^t_{\;\; \u t} d {\u t} = d {\u t}$ from Eq.(\ref{taufrw}).
Plugging this into Eq.(\ref{newgeo2}) one obtains $n=1$ and therefore has the solution\footnote{Note that 
this implies $\underline t^r = c_i$, and so $g_{\kappa \nu} \underline t^\kappa \underline t^\nu =0$, which
assures that $d\lambda$ is a constant of motion.}
\beqn
\underline\omega =  c_i ~a(t)  \quad, \quad  
{\u t}^{\u r}  =   c_i~ a(t)^2 \quad.  \label{antiomega} 
\eeqn
From this one defines the red shifts for both types of photons in the usual way as
\beqn
z = \frac{a_0}{a(t)} -1 \quad,\quad \underline z = \frac{a(t)}{a_0}  -1 \quad.  
\eeqn
Now let us examine Hubble's law. We make the expansions
\beqn
\frac{a(t)}{a_0} &=& 1 - H_0 \Delta_t - \frac{1}{2} q_0 H_0^2 \Delta_t^2 + {\cal O}(\Delta_t^3) \\
\frac{a_0}{a(t)} &=& 1 + H_0 \Delta_t -  H_0^2 \Delta_t^2 \left(1 + \frac{q_0}{2} \right) + {\cal O}(\Delta_t^3)  \quad,\\
\eeqn
with $\Delta_t = t_0 - t$ and the standard definitions
\beqn
H_0 = \frac{\dot a_0}{ a_0} \quad,\quad q_0 = - \frac{a_0 \ddot a_0}{\dot a_0^2}
\eeqn
One then finds to first order
\beqn
z = H_0~ \Delta_t \quad,\quad 
 {\u z} = - H_0~ \Delta_{t}  \quad. \label{hubble}
\eeqn
We see that the redshift goes into a blueshift and vice versa. Since we observe a redshift of
the light from far galaxies, two conclusions are possible: 
\begin{enumerate}
\item We are standard matter, then the universe is
currently expanding with $H_0>0$. 
\item Or we are anti-gravitating matter, then universe is currently shrinking $H_0<0$.
\end{enumerate}

\section{Cosmology}
\label{evolution}

To draw further conclusions, it is necessary to examine the evolution of the universe. 
We will formulate everything in the kinetic quantities, $\rho, p$ for the usual particles, and
$\underline \rho, \u p$ for the anti-gravitating particles. These quantities are positive in the common way. 
We define them in the comoving restframe and have as usual
\beqn
{\Theta}^{\nu\mu} &=& {\rm diag} (\rho,\frac{p}{a^2},\frac{p}{a^2 r^2},\frac{p}{a^2 r^2 \sin^2\theta}) \\
{\underline \Theta}^{\nu\mu} 
&=& {\rm diag} ({\underline\rho},\frac{\u p}{a^2},\frac{\u p}{a^2 r^2},\frac{\u p}{a^2 r^2 \sin^2\theta}) \label{uTheta} \quad,
\eeqn
and 
\beqn
{T}^{\nu\mu} 
&=& \Theta^{\nu\mu}  \\
{\underline T}^{\nu\mu} 
&=& {\rm diag} ({\underline \xi},\frac{\underline \chi}{a^2},\frac{\underline \chi}{a^2 r^2},\frac{\underline \chi}{a^2 r^2 \sin^2\theta}) 
\label{uT} \quad,
\eeqn
where the gravitational energy-density $\underline \xi$ and the gravitational pressure $\underline \chi$ fulfill an
unknown equation of state.

The {\sc SET} of usual fermionic matter (denoted with the upper index m) is identical to the kinetic {\sc SET}
\beqn
(T^{\rm m})^{\mu\nu} = (\Theta^{\rm m})^{\mu\nu}  = {\rm diag}(\rho
^{\rm m}, 0 , 0, 0 ) \quad,
\eeqn 
whereas for the anti-gravitating matter one has according to Eq. (\ref{sign}) 
\beqn
({\underline\Theta}^{\rm m})^{ \mu\nu} &=& 
{\rm diag} ( {\underline\rho}^{\rm m}, 0, 0, 0 ) \quad, \\  
({\underline T}^{\rm m})^{ \mu \nu} &=& -
{\rm diag}( {\underline\rho}^{\rm m}, 0, 0, 0 ) \quad.
\eeqn
For radiation (denoted with the upper index r), we use Belinfante's symmetrized kinetic {\sc SET}
\beqn
\left( {\Theta}^{\rm r} \right)^{\mu\nu} = \frac{1}{4 \pi} F^{\mu\lambda}F_{\lambda}^{\;\;\nu}
+ \frac{1}{16 \pi} g^{\mu\nu} F^{\kappa\lambda}F_{\kappa\lambda} \quad,
\eeqn
which is traceless $(\Theta^{\rm r})^\kappa_{\;\;\kappa} = 0$, and one concludes that also the
anti-gravitating matter has the ordinary equation of state for the kinetic quantities
\beqn
\underline\rho = 3 \u p \quad.
\eeqn
The anti-gravitating matter therefore does not behave like a Quintessence field. 

Further, it is insightful to examine the $t$-component of the conservation law 
for the kinetic {\sc SET}s Eq.({\ref{cons2}})
\beqn
\nabla_\nu \Theta^{\nu t} + \tau^t_{\;\;\underline\kappa} 
\nabla_\nu\underline\Theta^{\nu\underline\kappa} = 0 \quad. \label{const}
\eeqn
The first term yields the usual contribution
\beqn
\partial_t \rho + 3 \frac{\dot a}{a}\left(\rho+p\right)\quad.
\eeqn 
Since $\tau$ is diagonal and $\tau^t_{\;\;\underline t} = 1$, it is furthermore sufficient to evaluate the
expression
\beqn
\nabla_\nu \underline \Theta^{\nu \underline t} = \partial_t \underline\Theta^{t\underline t} 
+ \Gamma^{\underline t}_{\;\;\nu\underline\epsilon} \underline\Theta^{\nu\underline\epsilon} + 
\Gamma^\nu_{\;\;\nu\epsilon} \underline\Theta^{\epsilon \underline t}\quad. \label{constonlynew}
\eeqn
From Eq.(\ref{uTheta}) with Eq.(\ref{taufrw}) one has
\beqn
\underline \Theta^{\nu \underline \kappa} = \tau_\kappa^{\;\;\underline \kappa} \underline \Theta^{\nu\kappa} 
= {\rm diag}(\underline\rho,\u p, \u p,\u p) \quad.
\eeqn
Inserting the Christoffelsymbols from Appendix B, one finds that the second term in Eq.(\ref{constonlynew}) changes sign, and
Eq.(\ref{const}) takes the form
\beqn
\partial_t \left( \rho + \underline \rho \right) + 3\frac{\dot a}{a} \left( \rho + \underline 
\rho + p - \underline p \right) = 0 \quad. \label{consfrw}
\eeqn

Let us first examine the case of pure matter fields. Then, the above equation simplifies to
\beqn
\partial_t \left( \rho^{\rm m} + \underline \rho^{\rm m} \right) + 
3\frac{\dot a}{a} \left( \rho^{\rm m} + \underline \rho^{\rm m}\right) = 0 \quad. \label{consfrwm}
\eeqn
But one also has the Bianchi identities
\beqn
0 &=& \nabla_\nu \left( (\Theta ^{\rm m})^{\nu t} - (\underline \Theta^{\rm m})^{\nu t}\right) \nonumber \\
&=&  \partial_t \left( \rho^{\rm m} - \underline \rho^{\rm m} \right) + 3\frac{\dot a}{a} \left( \rho^{\rm m} - \underline \rho^{\rm m}\right) \quad.
\eeqn
These together with Eq.(\ref{consfrwm}) imply that both densities are separately conserved with the usual 
conservation law
 \beqn
0 &=& \partial_t  \rho^{\rm m}  + 3\frac{\dot a}{a}  \rho^{\rm m}  \\
0 &=& \partial_t  \underline \rho^{\rm m}   + 3\frac{\dot a}{a}   \underline \rho^{\rm m} \quad.
\eeqn
This means in particular that the energy density of anti-gravitating matter $\underline \rho^{\rm m}$ also dilutes 
with the volume $\sim 1/a^3$.

For radiation, the situation is slightly more complicated. Examination of Eq.(\ref{consfrw}) for purely anti-gravitating
radiation, $\rho^{\rm m} = \rho^{\rm r} = \underline \rho^{\rm m} =0$, yields
\beqn
0 = \partial_t \underline \rho^{\rm r} + 2 \frac{\dot a}{a}  \underline \rho^{\rm r}  \quad. 
\eeqn 
This is solved when $\underline \rho^{\rm r}$ dilutes with $\sim 1/a^2$. This fits very nicely with the
result of the previous section that the energy of anti-gravitating radiation experiences a blue shift instead
of a red shift. One then expects the energy density to scale like $E/V \sim a / a^3$. One can thus draw the 
important conclusion that the kinetic energy density of anti-gravitating radiation dilutes {\sl slower} than those
of anti-gravitating matter.  This scaling behavior can not directly be used for the cosmological evolution, since the
 quantities that enter the Friedmann-equations are the gravitational ones
$\underline \xi, \underline\chi$, whose properties remain to be investigated.
This however, allows us to discard the second possibility of the previous section. Apparently, the matter
we consist of is not radiation dominated. 
 
In general, one will be faced with a mixture of gravitating and anti-gravitating components.
To approach the problem, note that from Eq.(\ref{sign}) one has the relation
\beqn
({\underline T})^\mu_{\;\;\nu} + (\underline \Theta)^\mu_{\;\;\nu} =  
\delta^\mu_{\;\;\nu} \underline {\mathcal L} \quad,
\eeqn
or
\beqn
\underline \xi &=& - \underline \rho + \underline {\mathcal L} \\
\underline \chi &=& - \underline p - \underline {\mathcal L}
\eeqn
One can further use
\beqn
\partial_t \underline \xi + 3\frac{\dot a}{a} \left(\underline\xi +\underline \chi \right) = 0 \quad.
\eeqn
And rewrite this into  
\beqn
 \partial_t \underline \rho  + 3\frac{\dot a}{a} \left(\underline \rho +\underline p \right) =  \partial_t \underline 
{\mathcal L} \quad.
\eeqn 
In particular, for radiation one finds from this
\beqn
\partial_t \underline 
{\mathcal L}^{\rm r} = 2 \frac{\dot a}{a} \underline \rho^{\rm r}   \quad,   
\eeqn 
and thus ${\underline {\mathcal L}}^{\rm r} = - \underline \rho^{\rm r}$, and further $\underline \xi^{\rm r} 
= - 3 \underline \chi^{\rm r}$. Therefore, one can conclude that also the gravitational energy of the anti-gravitating radiation scales like $1/a^2$ and
  behaves like a curvature component. It consequently enters the Friedmann-equations as a curvature term does and
underlies similar experimental constraints. Inserting these results one obtains the modified 
Friedmann-Equations 
\beqn
\left( \frac{\ddot a}{a}\right)~ &=& - \frac{4 \pi G}{3} \left( \rho + 3 p 
+ \underline \rho - 3 \u p \right)  \label{1}\\
\left( \frac{\dot a}{a}\right)^2 &=&  \frac{8 \pi G}{3} \left( \rho - 2\underline \rho  \right)  \quad. \label{2}
\eeqn

\section{Early Universe}
\label{early}

To investigate the processes in the early universe, one has to keep in mind that the Lagrangian Eq.(\ref{full}) should
be interpreted as low-energy approximation of some underlying theory. At 
Planckian temperatures, higher order terms can become important, which will provide an interaction between 
gravitating and anti-gravitating matter fields. The strength of this interactions will typically grow 
with a power of the temperature over Planck mass $T/m_p$. Such  higher order interaction terms might e.g. 
take the form
\beqn
\sim \frac{1}{m_{\rm p}^2} \overline \Psi \slash D \tau_s^{-1} \slash D \underline \Psi \quad. 
\eeqn
However, the structure of these higher order terms, as well as
the particle content of the full theory are unknown. Therefore, we will aim to 
discuss the consequences on a general thermodynamical level by adding an exchange term $Q$ 
for the kinetic energies, which
becomes relevant at Planckian densities, and leaves the total kinetic energy conserved.

For such a coupled gravitating and anti-gravitating relativistic fluid, the 
conservation equations can then be written in the form  
\beqn
\partial_t \rho + 4 \frac{\dot a}{a} \rho &=& \frac{\dot a}{a} Q(\rho,\underline \rho) \quad,\label{consQ1}\\
\partial_t \underline\rho + 2 \frac{\dot a}{a}\underline\rho &=& - \frac{\dot a}{a} Q(\rho,\underline \rho)\label{consQ2} 
\quad.
\eeqn
The easiest choice for the exchange term is
\beqn
Q(\rho,\underline \rho) = \frac{\lambda}{m_{\rm p}^4} \rho \underline \rho \quad,
\eeqn
with some dimensionless constant $\lambda$ of order one. When the energy density of usual matter dominates that of the
anti-gravitating matter, then the interaction term should diminish it during a contraction with $\dot a<0$, 
and so it is sign$(\lambda)=$sign$(\rho-\underline \rho)>0$.  With this, the Eqs.
(\ref{consQ1},\ref{consQ2})
can be formally recast by attributing a time-dependent equation of state to the fields with 
\beqn
w &=& \frac{1}{3} - \frac{\lambda}{m_{\rm p}^4} \underline \rho \quad \mbox{for the usual, and} \label{w1}\\
\underline w &=& - \frac{1}{3} + \frac{\lambda}{m_{\rm p}^4} \rho \quad \mbox{for the anti-gravitating matter.}
\label{w2} 
\eeqn
The Friedmann equations then read (compare to Eqs.(\ref{1},\ref{2}))
\beqn
\left( \frac{\dot a}{a} \right)^2 &=& \frac{8\pi G}{3} \left( \rho - 2 \underline \rho \right)\\
\left( \frac{\ddot a}{a} \right) &=& - \frac{8\pi G}{3} \left( \rho - 3 Q(\rho,\underline \rho) \right) \quad.
\eeqn

These equations can be investigated qualitatively. Suppose an initial density distribution with 
$\rho_i-2\underline\rho_i=\tilde \rho_i >0$, and $\dot a>0$. In case the 
interaction term is negligible,  $\rho$ will drop faster than $\underline \rho$ and the 
expansion has a bounce when
$\rho = 2 \underline \rho$ is reached. In case the interaction term is not initially negligible, one sees from
Eq.(\ref{w2}) that $\underline \rho$ always dilutes, and so will the additional term 
in Eq.(\ref{w1}). 
Therefore, both correction terms in Eq.(\ref{w1},\ref{w2}) can eventually be neglected, and the previous case
applies. The anti-gravitating fluid acts then essentially like a positive curvature term, and 
a turning point is reached, at which the expansion stops.

From thereon, the universe goes into a contraction phase with $\dot a<0$. 
$\rho$ increases again, and does so
faster than $\underline \rho$. Both densities pass the initial configuration with a time-reversed $\dot a$.
 
When $\rho$ comes close to Planckian densities, one sees from Eq.(\ref{w2})  that its presence will
increase the $\underline w$ of the anti-gravitating matter, and $\underline \rho$ will eventually increase faster than $\rho$. When
$\underline\rho$ also reaches Planckian densities, it will furthermore slow the increase of $\rho$. That is, both
densities will again approach each other.

Eq.(\ref{consQ1}) shows that the interaction will re-distribute kinetic energy of the gravitational field into the
anti-gravitating matter, thereby increase $\partial_t \rho$ and decrease $\partial_t \underline \rho$. 
However, unlike the usual scenarios, the energy of the anti-gravitating field makes a negative gravitational contribution.
Therefore, $\ddot a$ can go from a positive to a negative values when $\underline \rho$ has increased to
\beqn
\underline \rho = \frac{m_{\rm p}^4}{3 \lambda} \quad.
\eeqn

Further approach of both densities results in reaching another turning point at Planckian densities, 
with $\rho = 2 \underline \rho$. From then on, the universe goes again into an expanding phase with 
positive acceleration,
until it reaches the initial configuration we started from (which might have had either positive or negative $\ddot a$, depending
on the initial values). Taken together, the evolution is a periodic cycle. The
maximal/minimal extension that can be reached in this cycle depends on the initial values of $\rho_i,\underline \rho_i$ 
and
the corresponding $a_i$.  
 
A special case occurs for the initial values $\rho_i = 2\underline \rho_i =2 m_{\rm p}^4 /(3 \lambda)$, for which all
quantities remain constant.

\begin{figure}[ht]
\vspace*{-0.1cm}
\hspace*{-0.5cm}\epsfig{figure=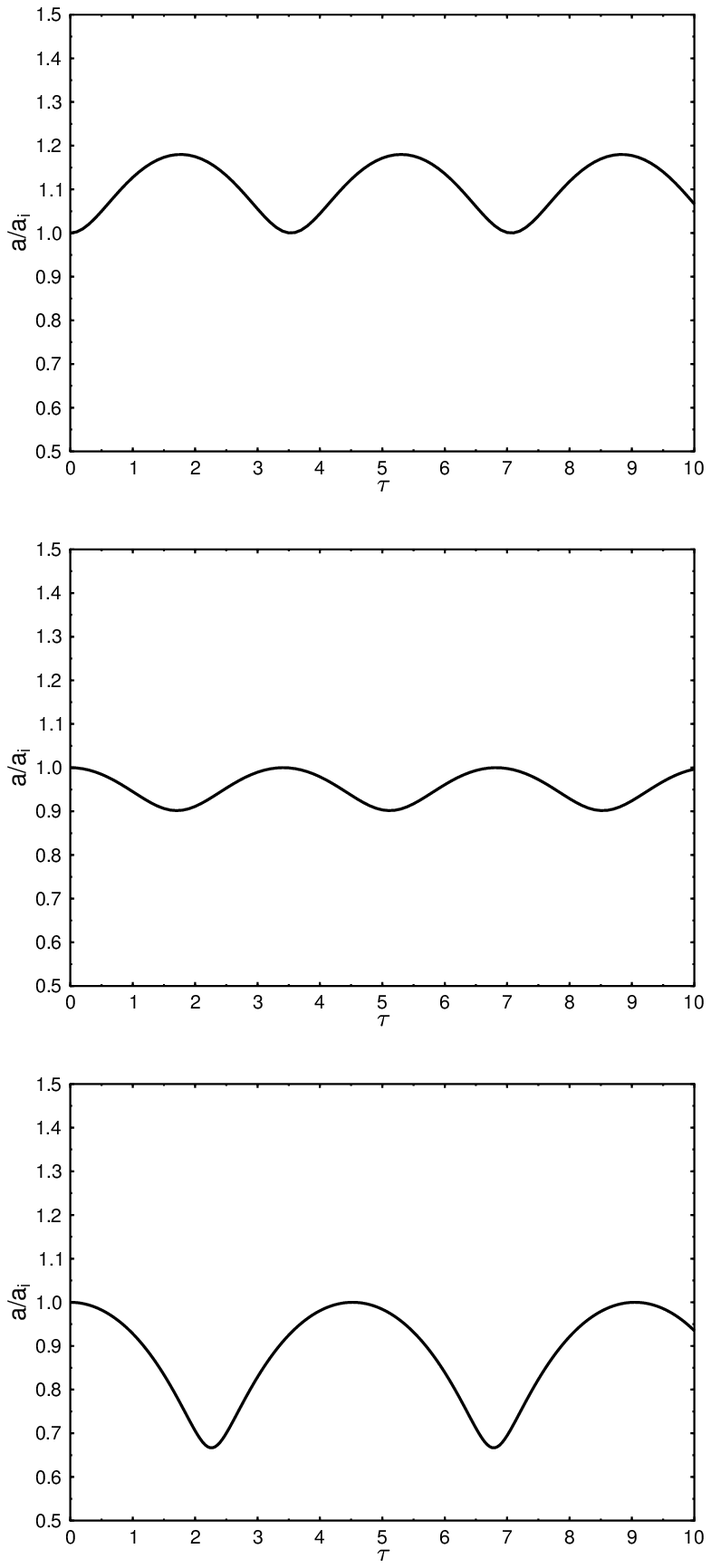, width=8.0cm}
\hspace*{-1.0cm}\epsfig{figure=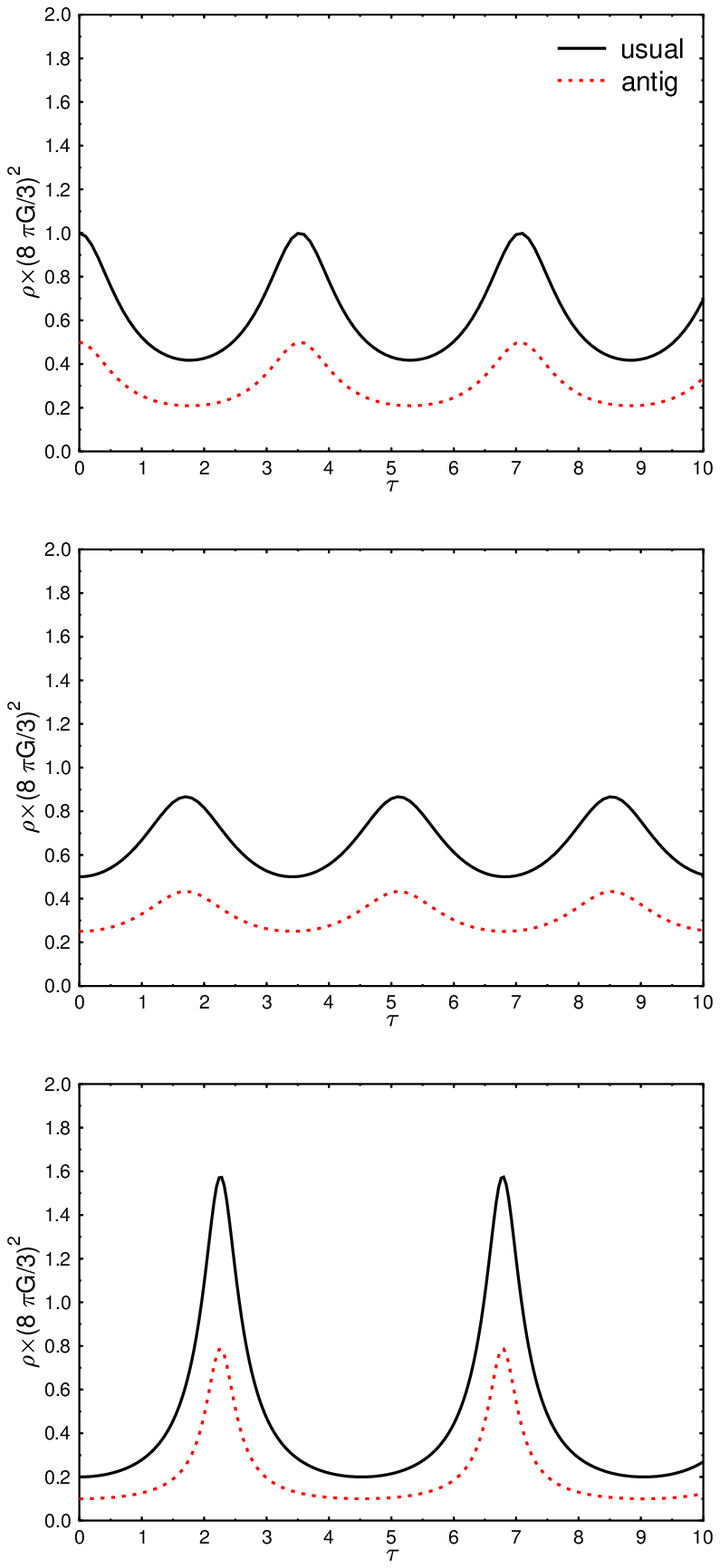, width=8.0cm}
\vspace*{-0.5cm}
\caption{Time evolution of the scale parameter $a$ (left), and of the energy densities (right) of usual (solid)
and anti-gravitating (dashed) matter. The horizontal axis shows  the dimensionless quantity $\tau = t \sqrt{8 \pi G/3}$. 
Here, the initial value $\dot a_i$ was set to zero, and $\lambda=1$. From the top to the bottom the
intial value of $\rho_i$ is 1, 0.5, 0.2 $\times (8 \pi G/3)^{-2}$. 
\label{fig3}}
\end{figure}

Figure \ref{fig3} shows some numerical results of the integration 
of these equations, with initial values $\rho_i$ of order $m_{\rm p}^4$, $\dot a_i =0$ and $\lambda=1$, 
which confirm this qualitative investigation.
One sees that for typical initial values with $\rho_i/\underline \rho_i$ of order $1-10$, the oscillation period is of order
a few Planck times. The larger the fraction of usual matter, the longer it will 
take until the anti-gravitational matter becomes important and the returning point it reached.

 %#######################################################################

\clearpage 
 
%#######################################################################

\section{Discussion}
\label{dis}

The fact that anti-gravitating radiation is the slowest to dilute, and that it should become
important only in the late stages of the universes expansion, requires that its initial density
is much smaller than that of usual matter. This problem is similar to the coincidence problem
in the context of a Cosmological Constant. 

However, an important fact to keep in mind is that an initially homogeneous distribution of the gravitating 
and anti-gravitating fluid is unstable under perturbations,
since the two kinds of matter want to separate. Small overdensities around the Planckian density of the 
discussed type will not lead to large deviations from the initial configuration, and the system will 
attempt to wash out perturbations through the strong interaction between both fluids. 
Large perturbations however, can lead to a drop of temperature and reaction rate, 
that lasts too long to enable return to the
initial configuration. Both types of matter can then separate into spatially 
distinct regions which predominantly consist of one type of matter. It is therefore natural to identify
the universe we currently experience as one such region with a dominantly positively gravitating matter
content.

Thus, the scenario investigated in the previous section applies for small perturbations in an 
initially Planckian density. For large perturbations, the evolution will attempt to break out of 
the oscillation at a maximum value of $a$.

To incorporate the two kinds of matter into the description one can treat the gravitating and
anti-gravitating matter as (spatially separated) homogenous distributions whose densities are time dependent.  
The anti-gravitating matter will be repelled from a positively gravitating fluid and be attracted by the
anti-gravitating one, and vice versa. Though the overall energy density is covariantly conserved, energy
flux will result in additional local source terms for both fluids, arising from the flux through the boundary.

Most importantly, the kinetic energy attributed to our universe is not necessarily conserved any longer.
In the process of repelling the anti-gravitational matter and attracting the usual matter, the universe experiences
an increase of the usual energy density. In particular, this increase will also
increase the acceleration relative to the standard cosmological model. 

Such a scenario can most intuitively be realized in a
higher dimensional space time, as e.g. examined in \cite{vandeBruck:2000ha}. In this case, our universe can be
described as a $3+1$ dimensional submanifold with dominantly gravitating matter, beneath which dominantly 
anti-gravitating sub-manifolds might be present. The scale parameter $a$ will not only be a function of $t$ but also a
function of the coordinates in the extra dimensions (in the simplest case one extra dimension). 
In the limit of infinitesimally thin branes, $a$ has to fulfill the common jump conditions 
associated with strongly localized layers of charges.  It would be interesting to further investigate 
the detailed properties of such a scenario. 

One should also note that the content of anti-gravitating matter in our universe 
will never vanish completely. Even though the
interaction between both types of matter is suppressed by the Planck scale, it will always take place. 
%The
%reaction rate is a power of $E \underline E / m_{\rm p}$, with the typical energies of the dominating
%matter components. Since the energy of a single anti-gravitating particle (radiation) grows with $a$, 
%whereas that of the usual particle (matter) remains constant, the reaction rate increases with the
%expansion of the universe. 

\section{Conclusions}
\label{concl}

We have studied the influence of anti-gravitating fields on the cosmological evolution. 
An examination of the geodesic equation for the anti-gravitating
photon in a Friedmann-Robertson-Walker background showed that this particle experiences 
a cosmological blue shift instead of a red shift. 

Furthermore, we have derived the modified evolution equation of the universe. These have additional
source terms arising from the anti-gravitating fields. Together with the stress-energy conservation,
these equations determine the evolution of the universe. We have shown that the anti-gravitating
matter density dilutes with the volume  of the universe $\sim 1/a^3$ and thus, like usually gravitating matter. 
The anti-gravitation radiation, however, dilutes due to its blueshift with $\sim 1/a^2$. In contrast
to Quintessence or a Cosmological Constant, it was shown that the anti-gravitating matter has a usual 
thermodynamical equation of state.

We have examined the evolution in the early universe with a strong interaction between 
both kinds of matter at Planckian densities. It was shown how such a scenario leads to a
periodic cycle which includes a phase of accelerated expansion. We also briefly discussed an extension
of the scenario, in which both types of matter separate on submanifolds in extra dimensions.

\section*{Acknowledgments}

 This work was supported by  the {\sc DFG} and by the Department of Energy under 
Contract DE-FG02-91ER40618. I thank the Perimeter Institute for kind hospitality.
%for helpful and stimulating discussions.

\begin{appendix}

\section{}

The Christoffelsymbols for the anti-gravitating particle in a Schwarzschild-metric can be computed using
Eq.(\ref{chrisag}). The coordinates are
$r,t,\theta,\phi$, and we use the notation $\gamma = 1 - 2M/r$.
\beqn
\Gamma^{ \underline t}_{\;\; t \underline r} &=& - \gamma \frac{M}{r^2} \quad,\quad
\Gamma^{ \underline t}_{\;\; r \underline t} = - \frac{1}{\gamma}   \frac{M}{r^2}  \quad,\quad
\Gamma^{ \underline r}_{\;\; t \underline t} = - \frac{1}{\gamma}  \frac{M}{r^2}   \nonumber \\
\Gamma^{ \underline r}_{\;\; r \underline r} &=&  \frac{1}{\gamma}  \frac{M}{r^2}  \quad,\quad
\Gamma^{ \underline r}_{\;\; \theta \underline\theta} =  \frac{1}{r}    \quad,\quad
\Gamma^{ \underline r}_{\;\; \phi \underline\phi} =  \frac{1}{r} \nonumber \\
\Gamma^{ \underline\theta}_{\;\; r \underline\theta} &=& - \frac{1}{r}    \quad,\quad
\Gamma^{ \underline\theta}_{\;\; \theta \underline r} =  - r \gamma  \quad,\quad
\Gamma^{ \underline\theta}_{\;\; \phi \underline\phi} =  \frac{\cos \theta}{\sin \theta} \nonumber \\
\Gamma^{ \underline\phi}_{\;\; r \underline\phi} &=&  -  \frac{1}{r} \quad,\quad 
\Gamma^{ \underline\phi}_{\;\; \theta \underline\phi} = - \frac{\cos \theta}{\sin \theta}    \quad,\quad
\Gamma^{ \underline\phi}_{\;\; \phi \underline r} =  - r \gamma \sin^2 \theta \nonumber \\
\Gamma^{ \underline\phi}_{\;\; \phi \underline\theta} &=& - \sin \theta \cos \theta   \quad.
\eeqn
All other entries are zero.

\section{}

The Christoffelsymbols for the gravitating and anti-gravitating particle in a FRW-background with metric element
Eq. (\ref{FRW}) are

\beqn
\Gamma^t_{\;rr} &=& a \dot a \quad,\quad
\Gamma^t_{\;\theta\theta} = a \dot a r^2 \quad,\quad \Gamma^t_{\;\phi\phi} = a \dot a r^2 \sin^2 \theta \nonumber\\
\Gamma^r_{\;tr} &=&   \Gamma^r_{\;rt} = 
\Gamma^\theta_{\;t\theta} = \Gamma^\theta_{\;\theta } =
\Gamma^\phi_{\;t\phi} =   \Gamma^\phi_{\;\phi t} = \frac{\dot a}{a}\nonumber\\
\Gamma^r_{\;\theta\theta} &=& -r \quad,\quad \Gamma^r_{\;\phi\phi} = -r \sin^2 \theta \nonumber\\
\Gamma^\theta_{\;r\theta} &=&  \Gamma^\theta_{\;\theta r} 
= \Gamma^\phi_{\;r\phi} =\Gamma^\phi_{\;\phi r} = \frac{1}{r} \nonumber\\
\Gamma^\theta_{\;\phi\phi} &=& - \sin \theta \cos \theta \quad,\quad
\Gamma^\phi_{\;\theta\phi}  = \Gamma^\phi_{\;\phi\theta} = \cot \theta  \label{chris}
\eeqn
And
\beqn
\Gamma^{ \underline t}_{\;\; r \underline r} &=& - \frac{\dot a}{a} \quad,\quad
\Gamma^{ \underline t}_{\;\; \theta \underline\theta} = - \frac{\dot a}{a}   \quad,\quad
\Gamma^{ \underline t}_{\;\; \phi \underline\phi} = - \frac{\dot a}{a} \nonumber \\
\Gamma^{ \underline r}_{\;\; t \underline r} &=&  - \frac{\dot a}{a}  \quad,\quad
\Gamma^{ \underline r}_{\;\; r \underline t} =  - a \dot a    \quad,\quad
\Gamma^{ \underline r}_{\;\; \theta \underline\theta} =  \frac{1}{r} \nonumber \\
\Gamma^{ \underline r}_{\;\; \phi \underline\phi} &=&  \frac{1}{r}    \quad,\quad
\Gamma^{ \underline\theta}_{\;\; t \underline\theta} =  - \frac{\dot a}{a} \quad,\quad
\Gamma^{ \underline\theta}_{\;\; r \underline\theta} =  - \frac{1}{r} \nonumber \\
\Gamma^{ \underline\theta}_{\;\; \theta \underline t} &=&  -  a r^2 \dot a \quad,\quad 
\Gamma^{ \underline\theta}_{\;\; \theta \underline r} = - r   \quad,\quad
\Gamma^{ \underline\theta}_{\;\; \phi \underline\phi} =  \frac{\cos\theta}{\sin \theta} \nonumber \\
\Gamma^{ \underline\phi}_{\;\; r \underline\phi} &=& - \frac{1}{r}  \quad,\quad 
\Gamma^{ \underline\phi}_{\;\; \theta \underline\phi} = - \frac{\cos\theta}{\sin\theta}   \quad,\quad
\Gamma^{ \underline\phi}_{\;\; t \underline\phi} =  - \frac{a}{\dot a} \nonumber \\
\Gamma^{ \underline\phi}_{\;\; \phi \underline t} &=& - a \dot a r^2 \sin^2\theta \quad,\quad 
\Gamma^{ \underline\phi}_{\;\; \phi \underline r} = - r \sin^2\theta   \quad,\quad
\Gamma^{ \underline\phi}_{\;\; \phi \underline\theta} =  - \sin\theta \cos \theta 
\eeqn
where $\dot{a} = \partial_t a$. All other entries are zero.

\end{appendix}

{ 
 
}
\end{document}